**Title:** Environmental Accounting for ecosystem conservation: Linking Societal and Ecosystem Metabolisms

**Authors:** Pedro L. Lomas[a,*], Mario Giampietro[a,b]

[a] Institut de Ciència i Tecnologia Ambientals (ICTA), Universitat Autònoma de Barcelona, 08193 Bellaterra, Spain.

[b] ICREA, Pg. Lluís Companys 23, 08010 Barcelona, Spain. *Corresponding autor: pedro.lomas@gmail.com

**Abstract:** This paper proposes an approach to environmental accounting useful for studying the feasibility of socio-economic systems in relation to the external constraints posed by ecological compatibility. The approach is based on a multi-scale analysis of the metabolic pattern of ecosystems and societies and it provides an integrated characterization of the resulting interaction. The text starts with a theoretical part explaining (i) the implicit epistemological revolution implied by the notion of ecosystem metabolism and the fund-flow model developed by Georgescu-Roegen applied to environmental accounting, and (ii) the potentials of this approach to create indicators to assess ecological integrity and environmental impacts. This revolution also makes it possible to carry out a multi-scale integrated assessment of ecosystem and societal metabolisms at the territorial level. In the second part, two applications of this approach using an indicator of the negentropic cost show the possibility to characterize in quantitative and qualitative terms degrees of alteration (crop cultivation, tree plantations) for different biomes (tropical and boreal forests). Also, a case study for land use scenarios has been included. The proposed approach represents an integrated multi-scale tool for the analysis of nature conservation scenarios and strategies.

**Highlights**

- The notions of social and ecosystem metabolisms allow to work at different scales
- Societal metabolism depends on the metabolic patterns of ecosystems embedding them
- Flow-fund model allow to link the social and ecosystem metabolism
- Metabolic pattern is used to define benchmarks for ecosystems to assess human impacts
- Genuine impact Indicators can be produced to make diagnosis and scenarios

# 1. Introduction

According to the results of the Millennium Ecosystem Assessment project, the increase in well-being experienced by part of the human population in the last 60 years has been achieved at the cost of the most extensive and rapid transformation of ecosystems in human history (Millennium Assessment, 2005). This explosion human activity on the planet has led some authors to propose the introduction of two concepts: (i) a new geological era called *Anthropocene* (Crutzen and Stoermer, 2000; Lewis and Maslin, 2015; Steffen et al., 2015a, 2011) to stress that currently biophysical processes controlled by humans represent the main driving force behind changes in the ecosystems (Zalasiewicz et al., 2008); and (ii) the notion of *planetary boundaries*, i.e., ecological limits for the human activity in order to operate safely within a global change framework (Rockström et al., 2009; Steffen et al., 2015b). The concept of planetary limits clashes with the economic strategy of perpetual growth, and implies acknowledging that the reproduction of the societal structures and functions depends on the integrity of ecological processes. In particular, two factors determine these limits to economic growth: its dependence on the availability of natural resources (limits of the supply capacity) and the damage that socio-economic activities implies on nature (limits of the sink capacity). For this reason, in the last decades there has been an increasing interest in developing approaches to improve the analysis of both the dependence and the impact of humans on ecosystems.

The ongoing effort to build an international framework on *environmental accounting* can be interpreted as a result of this interest (EEA, 2011; Obst, 2015; UN, 2014a, 2014b, 2014c; World Bank, 2010). This framework has the challenge to standardize the organization and presentation of useful information for characterizing the interface between the economy and the environment in order to support decision making (Vardon et al., 2016). In practical terms, this new System of Environmental Accounts is expected to complement the current System of National Accounts (UN, 2014a). This goal is approached by using two categories to define the elements describing socio-economic patterns in relation to nature: stocks of people and artefacts, and flows of energy and materials.

However, the ecosystem accounting framework developed continues to be labeled as "experimental", indicating that no complete agreement has been reached on how to carry out such a task (Bartelmus, 2015, 2014; UN, 2014b). The distinction proposed between stocks and flows have created many ambiguities when applied to multiple-scales analysis since the criteria used for defining these categories blur if non-equivalent descriptive domains and multiple boundaries are considered simultaneously (Giampietro and Lomas, 2014; Mayumi and Giampietro, 2014). Furthermore, the complex nature of the two systems analyzed implied that the methodologies proposed did not result completely satisfactory. Methodologies based on economic variables are in some cases effective in focusing on monetary benefits obtained by people exploiting ecosystems. However, they are not as effective in assessing the changes that this exploitation causes. On the other hand, methodologies based on biophysical indicators are effective in focusing on quantitative and qualitative changes suffered by ecosystems, but not as effective in assessing the consequences on the economy and the social well-being.

This dilemma points at a systemic conundrum of integrated assessment. To deal with this conundrum, it is very useful to frame the analysis of sustainability issues adopting the notion of *metabolism*. This concept assumes by default the co-existence of different relevant scales and dimensions of analysis (Giampietro, 2014; Giampietro et al., 2012). Thus, it becomes possible to characterize the reproduction of human societies by a continuous flow of energy and materials taken from and discarded to the environment, i.e. societal metabolism (Cottrell, 1955; White, 1943; Zipf, 1941). In the last decades, societal metabolism has been gaining momentum with the search for consistent environmental accounting methods for sustainability (Fischer-Kowalski, 1998a, 1998b, Giampietro, 2014, 1997, 1994; Giampietro et al., 1997; González de Molina and Toledo, 2014; Padovan, 2000).

An important contribution to this field has been provided by the *Bioeconomics* framework (Georgescu-Roegen, 1971; Giampietro et al., 2012; Mayumi, 2001). The bioeconomic framework moves the attention away from an input/output analysis of the various flows of goods and services consumed and produced to an analysis of *funds*, or the reproduction of production factors. This distinction between flows (inputs/outputs) and funds (estructural elements) makes it possible to explictitly address the issue of scale that appears when environmental boundary conditions are considered.

The aim of this paper is to present Multi-Scale Integrated Assessment of Societal and Ecosystem Metabolism (MuSIASEM) (Giampietro, 2004; Giampietro et al., 2013, 2012), based on the flow-fund model of Georgescu-Roegen, as an approach to make integrated assessments of society and nature. To this purpose, the theoretical basis for the concept of ecosystem metabolism, and the potentials of this approach to produce integrate assessments of the societal and ecosystem metabolisms are explained in section 2. To exemplify this potential, section 3 illustrates the reproduction of biomass as a fund, and section 4 presents three examples of application of this approach aimed at generating a quantitative assessment of the alteration level for terrestrial ecosystems: tree plantations, crop cultivation, and a hypothetical case study with different scenarios of land uses.

## 2. Theoretical basis

### 2.1. Ecosystem metabolism

Building on Lotka (1925), the ecologists E.P. Odum and H.T. Odum developed a methodological approach capable of generating quantitative analysis associated with the notion of *ecosystem metabolism* (Odum, 1957, 1956), becoming one of the most influential concepts in Systems ecology (Jørgensen, 2012). The general theoretical framework makes it possible a biophysical accounting of energy flows through networks, called *energy chains* (Odum, 1975). Energy chains define the relationships between different components making up an ecosystem, labeled as different *energy forms* (Odum, 1971; Odum and Odum, 1976). Ecosystems are represented in *energy flow diagrams* by using symbols carrying out specific meanings (Brown, 2004; Odum, 1994, 1983, 1971), as illustrated in figure 1.

The key feature of this approach is the possibility of integrating in quantitative terms information referring to different scales, whose identity is derived from the existing knowledge of ecosystems. This method makes it possible to establish a bridge between the metabolic characteristics of specific energy forms, observable at different levels of analysis. Thus, the characteristics of functional compartments (e.g. herbivores) at a meso level, can be linked to the characteristics of individual species of herbivores describing structural elements expressing the function (e.g. rabbits and dears) at a lower level of analysis. In the same way, functional elements representing the meso level can be linked to the characteristics of the whole ecosystem at the macro level. The ability of establishing these bridges across levels is important because the characteristics of emergent properties of the whole network (Odum, 1985, 1969; Odum et al., 1995) are observable only at the whole ecosystem level interacting with its context. This potential represents a remarkable feature of this accounting system, capable of handling the quantitative representation energy forms that are non-equivalent and non-reducible to each other using conventional mathematical models.

It must be noticed that this approach has been developed by using some of the most innovative scientific concepts of their time, in particular non-equilibrium thermodynamics applied to the ecological complex self-organizing systems (Glansdorff and Prigogine, 1971; Maturana and Varela, 1980; Nicolis and Prigogine, 1977; Schneider and Kay, 1994). The simultaneous adoption of thermodynamic and ecological narratives has some epistemological implications and assumptions. The thermodynamic narrative is used to describe the characteristics of the whole ecosystem, whose parts and functions are described using physical laws and conventional thermodynamic analysis. The ecological narrative is based on the assumption that biological and ecological processes of autopoiesis are taking place inside the system and are capable of stabilizing the identity of biological and ecological types at a local scale. It implies that the information stored in biological and ecological systems is reproduced and effectively used to maintain the expected characteristics of the functional and structural elements within the network, i.e. a given identity for metabolic elements, determining what should be considered as negative entropy for them. The concept of negative entropy (Schrödinger, 1967; page 78) makes it possible establish a link between the thermodynamic and ecological narrative. In fact, the definition given by Schrödinger refers to what is required from the environment by living (metabolic) systems. Thermodynamic constraints mandate a compatibility between internal processes of metabolism and the external processes determining the boundary conditions.

The simultaneous validity of these two narratives implies an impredicative relation (circular causality) between processes taking place at the same time at different scales: the metabolic characteristics of the parts (structural elements) determine the viability of the metabolic characteristics of the functional elements (bottom-up causation); and the metabolic characteristics associated with the required functions determine the feasibility of the metabolic characteristics of the structural elements (top-down causation). This is a well-known characteristic of complex systems organized over hierarchical levels (Giampietro, 1994; Pattee, 1973; Simon, 1962) called also holarchy (Koestler, 1969), double asymmetry (Grene, 1969) or equipollence (Iberall et al., 1980). The need to simultaneously describe metabolic processes at different space-time scales makes impossible to define a clear and unique boundary for the various elements, constraining the analyst to select a specific *environmental window of attention* (Odum, 1996, 1971). This fact forces the adoption of non-equivalent descriptive domains in which is impossible to reduce their representations of metrical and temporal relations to each other without losing relevant information (Giampietro et al., 2014, 2006a, 2006b). In particular, this multi-level characterization requires the integration of two views of the metabolic pattern of ecosystems (Giampietro, 2014, 2004; Giampietro et al., 2012), as shown in figure 1:

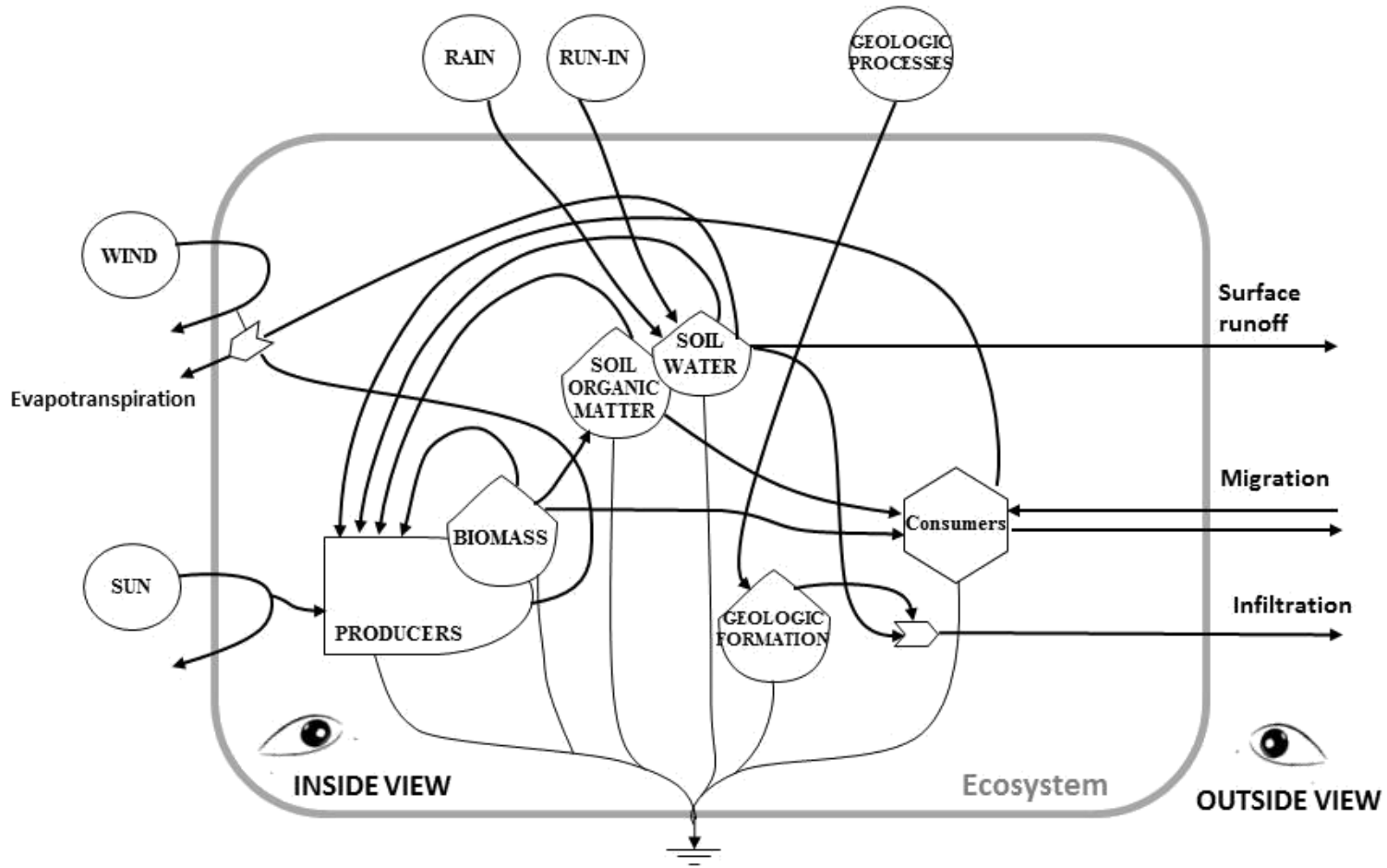

**FIGURE 1 ABOUT HERE**

*Outside view.* Based on the characterization of solar radiation, material (nutrients, organic matter), air, and water cycles supporting the biomass reproduction of the different ecosystem types. At this level, the application of the first law of thermodynamics and the principle of mass conservation make it possible to obtain the closure on the accounting.

*Inside view.* Associated to the specific choice of categories for the different energy forms considered within the graph. The inside view reflects the rationale of negative entropy. This implies that the characteristics of a metabolic element define what should be considered as an energy input for it. Within a metabolic network the definitions of energy forms are specific for the given elements belonging to the network.

Another important characteristic of this approach is the possibility of distinguishing expected *metabolic patterns* reflecting systemic properties of whole ecosystems (Odum, 1969). These patterns are determined by the required congruence between the relative size of the elements, their relative paces of energy dissipation (e.g., kg of meat eaten per kg of carnivore related to kg of plants per kg of carnivore), the topological relations over the flows across the nodes (e.g. the expected relation between functional compartments), and the compatibility of the whole network with boundary conditions (e.g. the available supply of inputs getting to the ecosystem). Later, analogous narratives about the idea of metabolic patterns have been developed (Allen and Hoekstra, 2015; Margalef, 1968, 1963, Ulanowicz, 1997, 1986).

The concept of ecosystem metabolic pattern is based on the hypothesis that after reaching a near closure of nutrient cycles, typical of mature ecosystems, informed autocatalytic loops or loop reinforcements (Odum, 1971) control the interactions taking place between a set of known energy forms. These interactions will tend to stabilize a given configuration of the network of transformations over attractors, determined by the starting point, boundary conditions, biological rules and physical laws. It should be noticed that the postulation of ecosystem typologies whose characteristics do not change significantly in time is a strong assumption. This ideal situation is almost impossible to find. However, in many cases it is possible to define expected metabolic characteristics through a set of benchmarks associated with specific ecosystem types. These benchmarks account for the quantity of inputs used or outputs formed per unit of ecosystem fund reproduced. The concepts of *Ecosystem Health* and *Ecological Integrity* (Crabbé et al., 2000; Karr, 1996; Kay et al., 2001; Kay and Regier, 2000; Pimentel et al., 2000; Waltner-Toews et al., 2008) were proposed using this idea.

## 2.2. Integrated assessment of societal and ecosystem metabolisms

Both society and ecosystems can be interpreted as complex, self-organizing, dissipative systems capable of stabilizing their own identity by reproducing a certain metabolic pattern. However, the mechanisms and processes involved in the expression of their respective interacting metabolic patterns operate at different scales or levels (figure 2).

**FIGURE 2 ABOUT HERE**

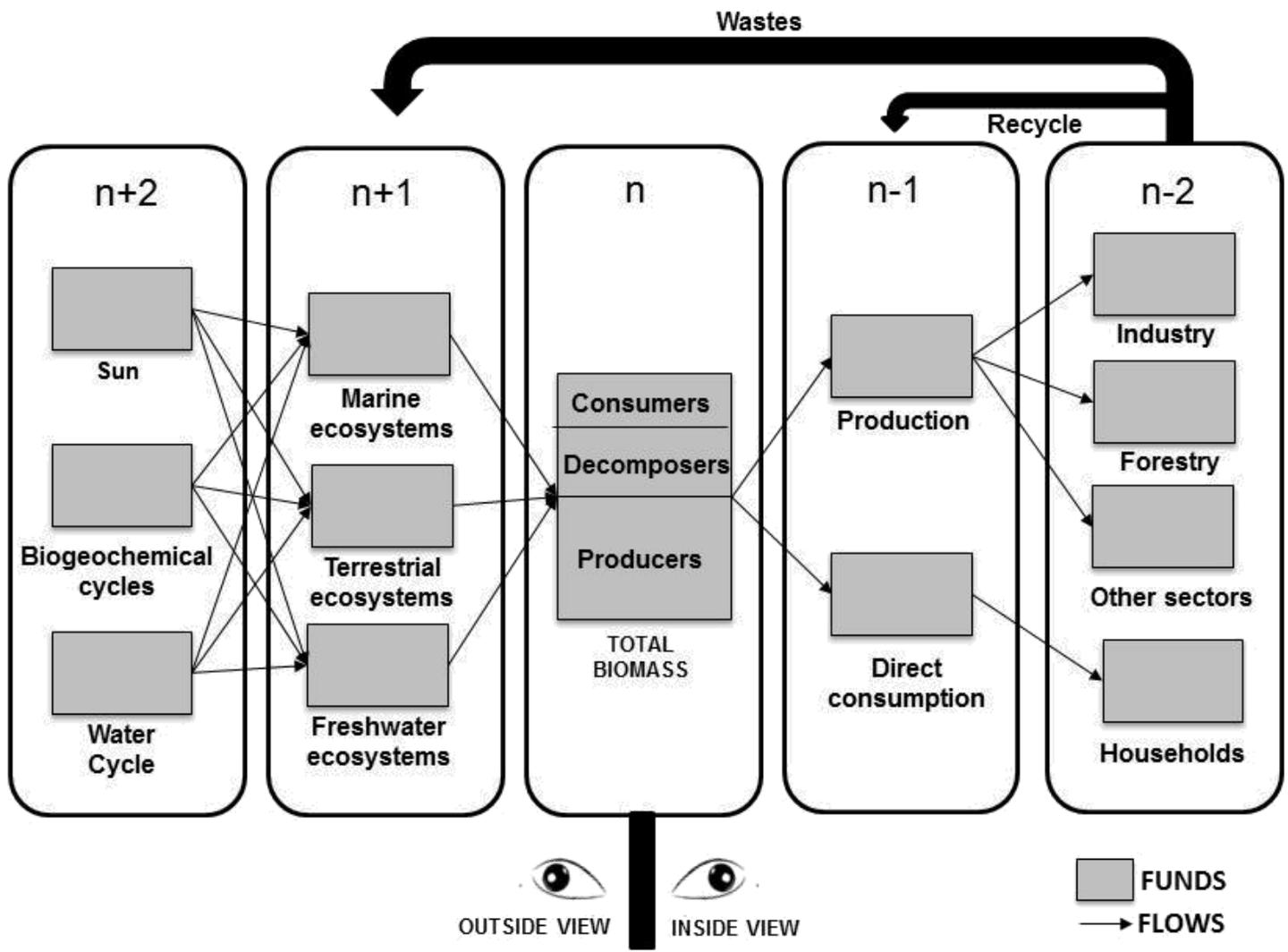

The metabolism of the aquatic, terrestrial and marine ecosystems (level n+1) is based on processes that are beyond direct human control, even if humans do affect these processes with their action. At a larger scale, the stabilization of biogeochemical cycles (level n+2) depends on a combination of natural processes of self-organization driven by thermodynamic laws and by the action of ecological funds, whose identity depends only in part of biological agents. By definition, autopoietic systems are expressing agency by articulating a direct control on the processes needed for their own reproduction. However, to be successful in this task they must rely on the existence of favorable boundary conditions that are guaranteed by processes outside their direct local control.

The same situation is found when looking at the societal metabolism. The control expressed by humans on socio-economic processes requires socio-economic *viability* (internal constrains) and social *desirability* (normative values) for the different economic and social sectors (levels n-1 and n-2), but also the pre-existence of favorable environmental boundary conditions or environmental *feasibility* (external constrains). These favorable conditions depend on resources made available by nature (level n) and by the stability of the metabolic patterns of ecosystems embedding societies (levels n+x).

The integrated assessment of this linked metabolism at different scales is the main goal of MuSIASEM. Under the MuSIASEM perspective, environmental accounting should work on the two main tasks previously defined in section 1: (i) characterizing the dependence of human activities on nature; and (ii) characterizing the impacts on ecological funds determined by human activities.

The dependence of human activities on nature can be assessed by a combination of quantitative and qualitative biophysical indicators at different levels. These indicators account for gross requirement of energy and materials from ecosystems and imports (direct and embodied) on the supply side, and the corresponding flow of wastes and exports (direct and embodied) on the sink side. Then, it becomes possible to calculate ratios describing the interaction society/environment, such as: the flow of resources from ecosystems needed per unit of societal fund reproduced or the flow of resources supplied to society per unit of ecosystem fund. These ratios can be used to study the effect of changes in societal or ecosystem metabolic patterns in real (diagnostic) or hypothetical (scenario) situations.

The unavoidable interaction of the two metabolisms implies the risk that an excessive growth in the activity of societal metabolism translates into an increasing damage to ecological funds and their reproduction, i.e., environmental impacts. Environmental impacts can be characterized: (i) on the supply side, in relation to the appropriation of energy and materials from ecosystems (overexploitation); and (ii) on the sink side, in relation to the ecosystems' absorption of wastes resulting from the economic processes (pollution). Ecological funds can also be indirectly harmed by an alteration of the ecological flows needed to support their reproduction. The impacts on ecological funds can be assessed by comparing the expected quantitative values characterizing states associated to ecological integrity with data coming from empirical studies describing the characteristics of altered ecosystems (or mosaics of land uses) which are supposed to have belonged originally to the given type. This method is at the basis of many ecological indicators such as e.g., *Environmental Loading Ratio* (Brown and Ulgiati, 1997; Odum, 1996; Ulgiati and Brown, 1998), the original framing of the *Ecological Footprint* (Wackernagel and Rees, 1998), *Ascendancy* (Ulanowicz, 1997, 2000), *Ecological Integrity Indexes* (Andreasen et al., 2001; Reza and Abdullah, 2011; Zampella et al., 2006), or *Nature Index* (Certain et al., 2011; Pedersen et al., 2013).

## 3. Methodology and calculations

In this example, standing biomass in terrestrial ecosystems has been treated as a fund to be reproduced (level n, in figure 2). It is known that the biomass reproduction (figure 3) is primarily associated to two important biogeochemical cycles: carbon and water (level n+2 in figure 2).

**FIGURE 3 ABOUT HERE**

The initial fund of biomass (SB) provides the ability to use the various inputs (solar energy, $CO_2$, water, nutrients) needed to reproduce itself through the solar energy captured via photosynthesis. The autotrophic compartment can generate the supply of energy that, when adopting the internal view, is seen as energy stored in the form of chemical bonds measured by gross primary productivity (GPP). Part of this GPP is used by the plants itself in the autotrophic respiration ($R_a$) and therefore is not available to the rest of the compartments. The fraction of energy which is not consumed in this internal loop is available for the synthesis of plant tissues (foliage, wood and roots) and constitutes the net primary productivity (NPP). NPP is then used either to increase the capital (+ΔSB, an increase in the standing biomass), largely in non-mature ecosystems, or to sustain the heterotrophic respiration ($R_h$), fueling the activity of the other compartments within the ecosystem, mostly in mature ecosystems.

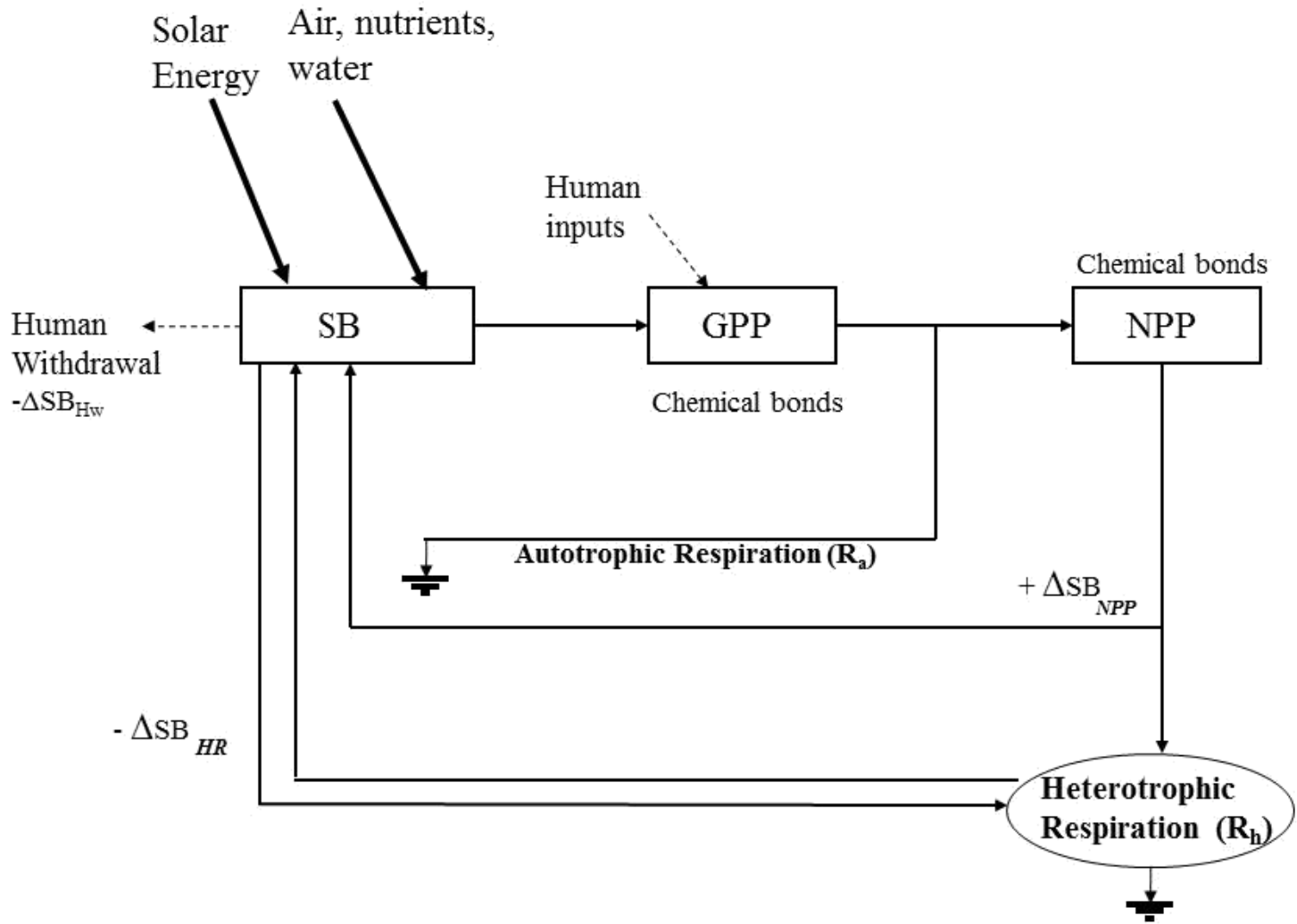

According to this set of relations, within each typology of terrestrial ecosystems (level n+1 in figure 2) it is expected to find some general features. Under thermodynamics principles, the undisturbed state for a terrestrial ecosystem implies the simultaneous minimization of the negentropic cost (Prigogine, 1955) and maximization of energy dissipation (Lotka, 1925; Odum and Pinkerton, 1955). This expected pattern can be used to develop indicators useful to characterize ecosystem integrity and to detect human alteration (level n-1 in figure 2). The level of alteration is flagged by a disturbance in this set of expected relations moving the characteristics of the disturbed ecosystem in the reverse direction.

In this context, the thermodynamic cost or *negentropic cost* ($\Phi$), defined as the amount of energy required to support the metabolic processes per unit of standing biomass (Giampietro et al., 1992; Giampietro and Pimentel, 1991), is one example of flow-fund ratio that can be used to express the alteration of terrestrial ecosystems. The alteration can be measured by looking at a decrease in SB (absolute reduction of the fund) and/or at the ratio of energy metabolized per unit of SB maintained (increase in the fund cost) (Giampietro, 1999).

Since water is needed as a direct input to photosynthesis, to maintain tissue functionality, to transport nutrients from the roots to the leaves, and to cool and maintain the turgidity of vegetative structures, this flow-fund ratio can be characterized in terms of the energy associated to the required flow of water evapotranspiration (ET), i.e., the thermal dissipation associated with the maintenance of SB. But an assessment of ET is not easy, since it implies a mix of physical (evaporation) and biochemical (transpiration) processes that are difficult to separate and assess. Usually, the amount of water lost during the production of biomass or the fixation of $CO_2$ in photosynthesis is known as transpiration efficiency (TE) or water-use efficiency (WUE), when talking about the biomass produced per unit of water transpiration (Lambers et al., 2008).

In this way, it becomes possible to calculate the energy dissipation (W/m$^2$) of transpiration associated to production per unit of SB (kg/m$^2$) in terrestrial biomes, and obtain a benchmark value of negentropic cost (W/kg) that can be used as an external reference to be compared with actual systems and to link this assessment to spatial analysis. The procedure is explained in Eq1:

$$\Phi = \frac{WF}{SB} = \frac{GPP * TE}{SB} = \frac{GPP * \frac{1}{WUE}}{SB} \quad (Eq\ 1)$$

$\Phi$ = negentropic cost (W/kg), WF=energy dissipation of transpiration associated to production (W/m$^2$); SB=standing biomass (kg of biomass/m$^2$), GPP=Gross Primary Production (kg of biomass produced/m$^2$*yr); WUE=Water-use efficiency (kg of biomass produced /kg of water transpired); TE =Transpiration efficiency =1/WUE (kg of water transpired /kg of biomass produced).

By using this set of quantitative relations, it becomes possible, at a given scale, to compare the performance of different land uses with the performance of natural systems substituted by them. A combination of indicators can be used to quantify not only pressures, assessing inflows or outflows, but also impacts, assessing differences in the quantity of inflows or outflows needed per unit of fund sustained. In the next section, the use of this approach is illustrated in two examples.

# 4. Results and discussion

## 4.1. Comparing the performance of different typologies: Cultivation in tropical lands

Figure 4 is a scatterplot presenting data from literature about the total quantity of fund (SB) per unit of area respect to the negentropic cost of fund units for some mature tropical rainforests (n= 9) (Clark et al., 2013) and some selected monoculture crops from FAOSTAT (n=9) frequently cultivated in different countries (Navarro et al., 2008), representing both paleotropical and neotropical conditions.

**FIGURE 4 ABOUT HERE**

In this graph, the values of flow/fund ratio for the non-dominated tropical forests can be considered as benchmarks for the natural ecosystems metabolic pattern, i.e., the ecosystem integrity external reference for this typology of ecosystem. Benchmark values of negentropic cost for mature tropical forests in the scatterplot are within the range of 0.4-1.8 W/kg of SB. Whereas, the data describing the metabolic pattern of crops indicate the situation for the altered state. Negentropic values of crops cultivated under tropical conditions represented in the scatterplot are within the range of 13-57 W/kg of SB.

Thus, within this representation of the metabolic pattern, in the tropics, the assessment of the average value for the flow-fund negentropic cost ratio calculated to croplands (Mean=

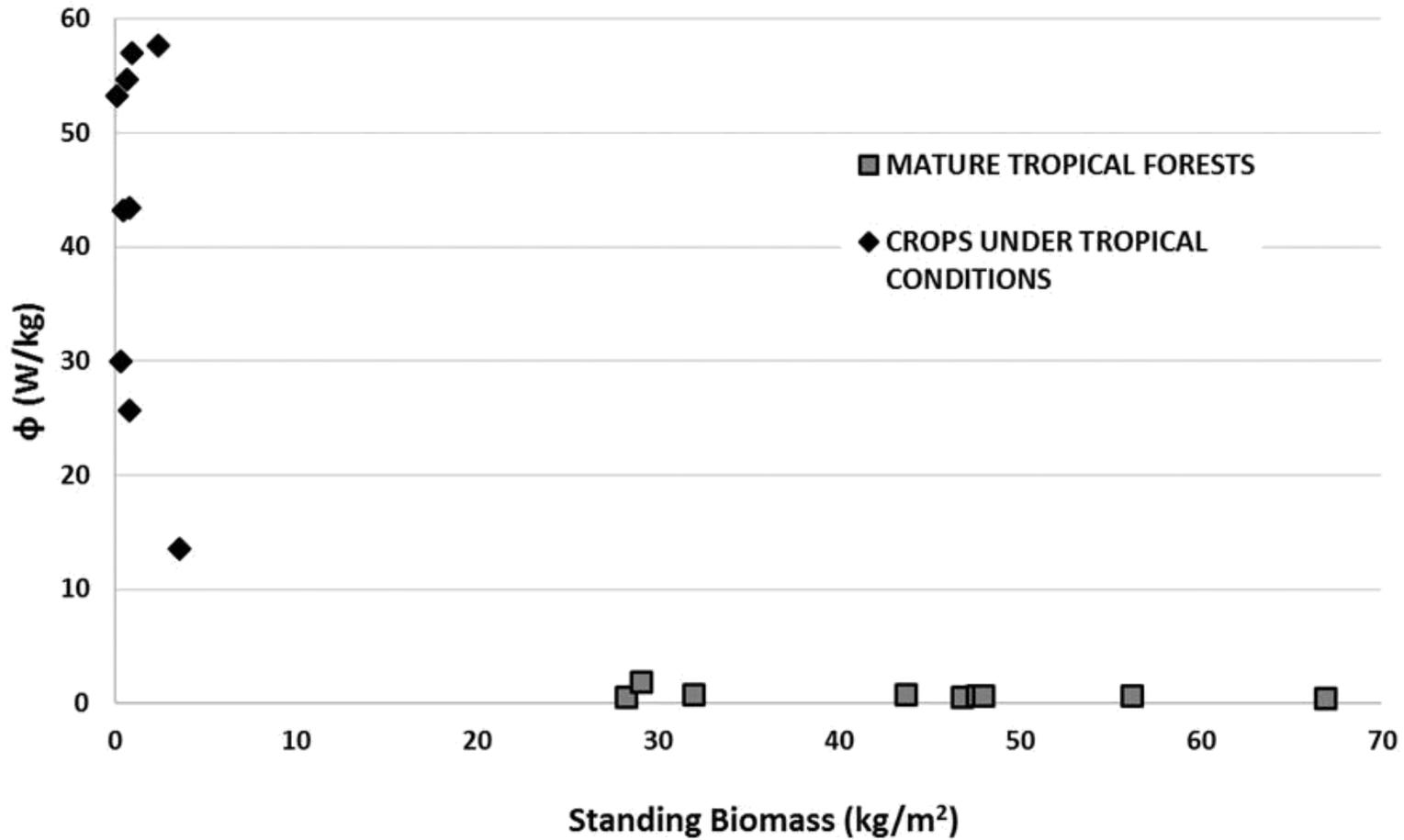

42 W/kg, SD =15.7) is around 60 times larger than the average negentropic cost value for the original tropical rainforest biomass (Mean= 0.7 W/kg, SD= 0.4).

With monoculture crops, largely used in the tropics, agriculture practices tend to reduce the SB on the territory. According to the sample data used in this case study, average SB in mature tropical rainforests (Mean= 44.3 kg/m$^2$, SD=12.9) is around 44 times larger than the average SB for crops cultivated under tropical conditions (Mean= 1.1 kg/m$^2$, SD=1.1). In practice, substitution of original tropical forests by monoculture crops means reducing the reproduction of ecological funds capable of using solar energy, recycling nutrients, and optimizing the use of water to produce biomass (Nicholls et al., 2016).

Consequently, the mode of production based on High External Input Agriculture (HEIA) can be characterized in this analysis as a stock-flow strategy boosting primary productivity (the flow of biomass taken away) for human withdrawal (crop yields) through the continuous increase in the required investment of societal funds (human labor and technology or power capacity) and flows (pesticides and herbicides, fertilizers, fuels) under human control in order to compensate the elimination of ecological processes, and regulate the rate of extraction.

### 4.2. Comparing the performance of different typologies: Forest plantations under boreal conditions

Following the proverb *trees alone do not make a forest*, the same type of difference between ecosystem integrity and altered state can be found when comparing the metabolic patterns of non-altered forest ecosystems with forest plantations located in territories previously covered by this typology of terrestrial ecosystem.

Analogous to the previous case study, figure 5 provides another scatterplot presenting data from literature about the total quantity of fund (standing biomass) per unit of area respect to the SB negentropic cost for some Class I mature boreal forests (n=23) around the world (Gower et al., 2012) and selected tree plantations under boreal conditions (n=61) in the north-east and south-west regions of China (Zhao and Zhou, 2005).

**FIGURE 5 ABOUT HERE**

As shown in the figure 5, the natural metabolic pattern associated with the ecological integrity of these boreal forests is characterized by benchmark values of SB negentropic cost within the range of 0.3-1.9 W/kg of SB; whereas the estimations made about forest plantations in boreal areas of China (the altered state used for this example) include values of negentropic cost in the range of 6.8-26.9 W/kg of SB.

Thus, the development of forest plantations for commercial purposes under boreal conditions presents a negentropic cost (Mean= 13.7, SD = 5.1) around 10 times larger than the negentropic cost associated to the mature forests originally established in these kind of areas (Mean= 1.4, SD= 1.8). According to this indicator, the differentiation between forests and tree plantations is plainly clear.

Analogous to the alteration associated to cultivation in tropical crops, the SB negentropic cost associated to tree plantations is higher than the one linked to mature boreal forests.

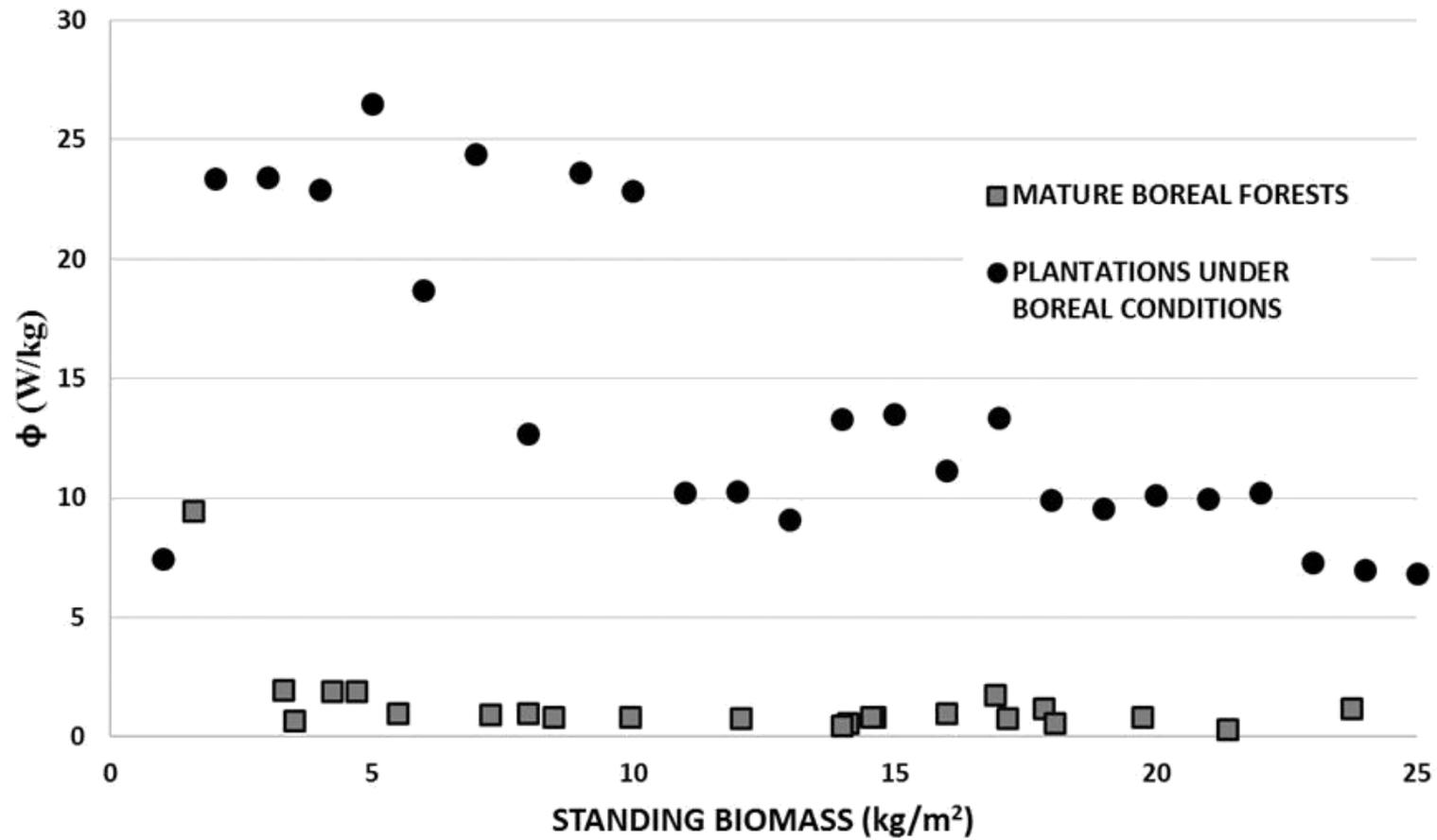

But in this case, average SB in mature boreal forests (Mean = 12 kg/m$^2$, SD= 6.4) is only 1.4 times larger than the average SB for tree plantations (Mean = 8.5 kg/m$^2$, SD=5.4), probably showing the variability on their stand ages and environmental history. Therefore, longstanding tree plantations (> 40 years) could present similar SB patterns to some young mature forests (< 40 years) (Xie et al., 2011). Here, tree plantations present a faster turnover rate and a higher negentropic cost due to commercial purposes; whereas, the natural boreal forest, without human interference, increases the standing biomass at a lower turnover rate to fit the potential determined by the biogeophysical conditions with a lower negentropic cost.

This mode of production can be characterized as a rapid fund-flow or slow stock-flow strategies, in which old tree plantations can reach acceptable levels of SB before being felled but a turnover rate faster than the natural boreal forests substituted.

However, the difference between the negentropic ratios of the altered and natural states is smaller in the case of tree plantations than the difference found for crop cultivation and their respective substituted forests. Under this type of analysis, tree plantations for commercial purposes result to be a less altered form of terrestrial ecosystem than crop monocultures if ecological integrity characterized by benchmark values of the terrestrial ecosystems being substituted is considered.

### 4.3. From typologies to individuals: Simulation of scenarios for conservation purposes

In real cases, a land use mosaic or potential scenarios for different land uses need to be assessed in a specific territory. Thus, not only intensive indicators of performance for different ecosystem or land use typologies are relevant but also individual size, i.e., extensive indicators quantifying the actual size of the characteristic considered for the different typologies performing at the scale studied.

As explained elsewhere (Aspinall and Serrano-Tovar, 2014; Serrano-Tovar and Giampietro, 2014), this approach can also be used to go beyond the comparison of performances for different ecosystem and land use typologies through the use of conventional GIS tools.

In this example, values of energy required to support the metabolic processes of certain ecosystem or land use typology per area unit (calculated in $W/m^2$ in these study cases) can be obtained by using benchmark values (expressed in W/kg of biomass for these examples) linked to their respective funds (in these examples, in terms of kg of biomass/$m^2$). Calculating the area for any of the land use or ecosystem patches and multiplying it by the energy requirements per area unit, different scenarios for land use arrangements or activities can be assessed in terms of negentropic cost (Eq2) in order to support decision-making processes with information about ecological integrity:

$$\Phi_{total} = \Phi_1 * Area_1 + \Phi_2 * Area_2 \ldots + \Phi_n * Area_n \quad (Eq2)$$

, where $\Phi_{total}$=total negentropic cost for the scenario proposed (W), $\Phi_x$=total negentropic cost per area unit for land use x ($W/m^2$), $Area_x$= total land use x area ($m^2$).

Figure 6 illustrates a hypothetical case study based on the data obtained to the boreal forests in China, already presented in the section 4.2.

**FIGURE 6 ABOUT HERE**

On the left side, it is presented the current state scenario for the case study of land use with a 75 % of mature boreal forest (*Larix gmelinii*) with a negentropic cost ($\Phi$) about 6.4 W/m$^2$, and 25 % of tree plantations (*Larix*) with an associated negentropic cost ($\Phi$) of 57.4 W/m$^2$. Negentropic cost ($\Phi$) for 100,000 m$^2$ of total area in this hypothetical case study is 1.9 MW.

Two scenarios of development have been proposed. The scenario 1, in which the extension of *Larix* plantations grows to reach 45 % of total area, and the scenario 2, in which total extension of tree plantations also reaches 45 %, but *Larix* plantations only reach 30 % of the total area and the rest of the area dedicated to tree plantations is occupied by *Pinus* with a negentropic cost ($\Phi$) of 106.3 W/m$^2$. Thus, total negentropic costs ($\Phi$) for 100,000 m$^2$ of total area in the scenarios 1 and 2 are 2.9 MW and 3.7 MW, respectively.

## 5. Conclusions

Several issues derived from the multiplicity of scales involved and the different (non-equivalent) descriptive domains involved make it difficult to find effective methods to account for the environmental impacts associated to socio-economic activities and its dependence on ecosystems. In this article, an application of the MuSIASEM approach,

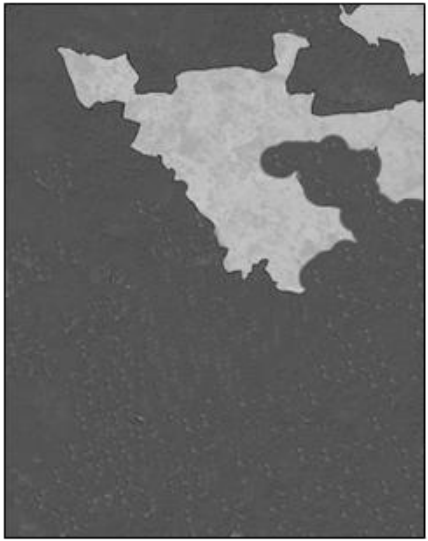
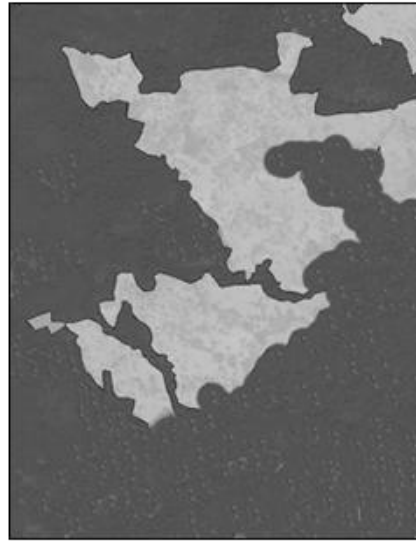
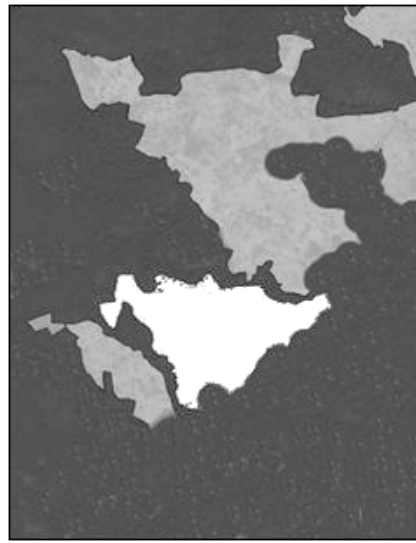

**CURRENT STATE**
75 % Mature forest (Φ=6.4 W/m²)
25 % Plantation *Larix* (Φ=57.4 W/m²)

Φ = 1.9 MW

**SCENARIO 1**
LAND USE
55 % Mature forest (Φ=6.4 W/m²)
45 % Plantation *Larix* (Φ=57.4 W/m²)

Φ = 2.9 MW

**SCENARIO 2**
LAND USE
55 % Mature forest (Φ=6.4 W/m²)
30 % Plantation *Larix* (Φ=57.4 W/m²)
15 % Plantation *Pinus* (Φ=106.3 W/m²)

Φ = 3.7 MW

MATURE FOREST
PLANTATION *Pinus*
PLANTATION *Larix*

Total Area = 100000 m²

based on the conceptual flow-fund model and the notion of metabolism, has been used to deal with this challenge.

By leaving the simplistic distinction between stock and flows, and moving to a more complex categorization based on fund-flows and stock-flows, it becomes possible to better frame and quantify human-nature interactions at different scales. In particular, it is possible to characterize both the contribution of ecosystems to human activities and the level of alteration of ecosystem funds when considering the throughputs of materials and energy metabolized by society.

In this analytical framework, society and ecosystems are interpreted as complex, self-organizing, dissipative systems capable of stabilizing their own identity by reproducing a certain metabolic pattern at different scales: the societal and the ecosystem metabolisms, respectively. This representation can be used to characterize ecological integrity for different typologies of un-altered ecosystems by means of benchmark values used to assess the stress levels induced by human exploitation. At the same time, societal metabolic patterns depend on the stability of the ecosystems embedding them. Thus, sustainability can be defined in terms of social desirability, socio-economic viability and ecological feasibility of human activities. The information on both the characteristics of societal and ecosystem elements at different scales makes it possible to develop a diagnostic and a simulator tool kit in relation to the study of ecological feasibility for conservation purposes.

Some examples of how this approach can be applied to develop a diagnostic tool for the analysis of alteration in terrestrial ecosystems have been presented. Using benchmark values, the status of two types of managed terrestrial ecosystems (tree plantations in boreal areas and crop cultivation in tropical zones) has been compared to the expected status of the corresponding un-altered ecosystems of reference. In relation to the task of environmental accounting, the external constraints limiting the amount of biomass used by societal metabolism have been studied. They were determined by available land, net supply per hectare and level of stress that the chosen form of exploitation implies on the ecological system,

Concerning the use of the approach to generate scenarios, the example given has illustrated the status of two different patterns of land use for tree plantations under boreal conditions. In this way, the impact of different land use policies on ecological integrity can be studied. The relative area for the different types of ecosystems in the considered pattern of land use can be associated to different levels of impact for different forms of exploitation.

**Acknowledgements**

Pedro L. Lomas wants to thank the Alliance 4 Universities for his post-doctoral grant. Authors want to thank the two anonymous reviewers for their useful comments to improve the early version of this paper. Part of the work of Mario Giampietro has received funding from the European Union's Horizon 2020 research and innovation programme under grant agreement No. 689669 (MAGIC). This work reflects only the authors' view and the funding agencies are not responsible for any use that may be made of the information it contains.

# References

Allen, T.F.H., Hoekstra, T.W., 2015. Toward a unified ecology. Columbia University Press, New York, USA.

Andreasen, J.K., O'Neill, R. V, Noss, R., Slosser, N.C., 2001. Considerations for the development of a terrestrial index of ecological integrity. Ecol. Indic. 1, 21–35.

Aspinall, R., Serrano-Tovar, T., 2014. GIS protocols for use with MuSIASEM, in: Giampietro, M., Aspinall, R.J., Ramos-Martin, J., Bukkens, S.G.F. (Eds.), Resource Accounting for Sustainability: The Nexus between Energy, Food, Water and Land Use. Routledge, London, UK, pp. 135–146.

Bartelmus, P., 2015. Do we need ecosystem accounts? Ecol. Econ. 118, 292–298.

Bartelmus, P., 2014. Environmental-Economic Accounting: Progress and Digression in the SEEA Revisions. Rev. Income Wealth 60, 887–904.

Brown, M.T., 2004. A picture is worth a thousand words: energy systems language and simulation. Ecol. Modell. 178, 83–100.

Brown, M.T., Ulgiati, S., 1997. Emergy-based indices and ratios to evaluate sustainability: monitoring economies and technology toward environmentally sound innovation. Ecol. Eng. 9, 51–69.

Certain, G., Skarpaas, O., Bjerke, J.-W., Framstad, E., Lindholm, M., Nilsen, J.-E., Norderhaug, A., Oug, E., Pedersen, H.-C., Schartau, A.-K., van der Meeren, G.I., Aslaksen, I., Engen, S., Garnåsjordet, P.-A., Kvaløy, P., Lillegård, M., Yoccoz, N.G., Nybø, S., 2011. The Nature Index: a general framework for synthesizing knowledge on the state of biodiversity. PLoS One 6, e18930.

Clark, D.A., Brown, S., Kicklighter, D.W., Chambers, J.Q., Thomlinson, J.R., Ni, J., Holland, E.A., 2013. NPP Tropical Forest: Consistent Worldwide Site Estimates, 1967-1999, R1. Data set. Available on-line [http://daac.ornl.gov]. doi:10.3334/ORNLDAAC/616

Cottrell, F., 1955. Energy and society: the relation between energy, social change, and economic development. Greenwood Press, Westport Connecticut.

Crabbé, P., Holland, A., Ryszkowki, L., Westra, L. (Eds.), 2000. Implementing ecological integrity: Restoring regional and global environmental and human health. Kluwer Academic Publishers, Dordrecht, The Netherlands.

Crutzen, P.J., Stoermer, E.F., 2000. The Anthropocene. IGBP Newsl. 41, 17–18.

EEA, 2011. An experimental framework for ecosystem capital accounting in Europe. Copenhagen, Denmark.

Fischer-Kowalski, M., 1998a. Society's metabolism. The Intellectual history of Materials Flow Analysis, Part II, 1970-1998. J. Ind. Ecol. 2, 107–136.

Fischer-Kowalski, M., 1998b. Society's metabolism. The Intellectual history of Materials Flow Analysis, Part I, 1860-1970. J. Ind. Ecol. 2, 61–78.

Georgescu-Roegen, N., 1971. The entropy law and the economic process. Harvard University Press, Cambridge, MA, USA.

Giampietro, M., 2014. Scientific basis of the narrative of metabolism, in: Giampietro, M., Aspinall, R.J., Ramos-Martin, J., Bukkens, S.G.F. (Eds.), Resource Accounting for Sustainability: The Nexus between Energy, Food, Water and Land Use. Routledge, London, UK, pp. 22–32.

Giampietro, M., 2004. Multi-Scale Integrated Analysis of Agroecosystems. CRC Press, Boca Raton, FL, USA.

Giampietro, M., 1999. Economic growth, human disturbance to ecological systems and sustainability., in: Walker, L.R. (Ed.), Ecosystems of Disturbed Ground, Ecosystems of the World. Elsevier, Amsterdam, The Netherlands, pp. 741–763.

Giampietro, M., 1997. The link between resources, technology and standard of living: A theoretical model, in: Freese, L. (Ed.), Advances in Human Ecology. JAI Press, Greenwich, CT, USA, pp. 73–128.

Giampietro, M., 1994. Using hierarchy theory to explore the concept of sustainable development. Futures 26, 616–625.

Giampietro, M., Allen, T.F.H., Mayumi, K., 2006a. Science for governance: the implications of the complexity revolution, in: Guimaraes-Pereira, A., Guedes-Vaz, S., Tognetti, S. (Eds.), Interfaces between Science and Society. Greenleft Publishing, Sheffield, UK., pp. 82–99.

Giampietro, M., Allen, T.F.H., Mayumi, K., 2006b. The epistemological predicament associated with purposive quantitative analysis: Complexity and Ecological Economics. Ecol. Complex. 3, 307–327.

Giampietro, M., Aspinall, R.J., Ramos-Martín, J., Bukkens, S.G.F., 2014. Resource Accounting for Sustainability: The Nexus Between Energy, Food, Water and Land Use. Routledge, London, UK.

Giampietro, M., Bukkens, S.G.F., Pimentel, D., 1997. The link between resources, technology and standard of living: Examples and applications, in: Freese, L. (Ed.), Advances in Human Ecology. JAI Press, Greenwich, CT, USA, pp. 129–199.

Giampietro, M., Cerretelli, G., Pimentel, D., 1992. Energy analysis of agricultural ecosystem management: human return and sustainability. Agric. Ecosyst. Environ. 38, 219–244.

Giampietro, M., Lomas, P.L., 2014. The interface between societal and ecosystem metabolism., in: Giampietro, M., Aspinall, R.J., Ramos-Martin, J., Bukkens, S.G.F. (Eds.), Resource Accounting for Sustainability: The Nexus between Energy, Food, Water and Land Use. Routledge, London, UK.

Giampietro, M., Mayumi, K., Sorman, A.H., 2013. Energy Analysis for a Sustainable Future: Multi-Scale Integrated Analysis of Societal and Ecosystem Metabolism. Routledge, London, UK.

Giampietro, M., Mayumi, K., Sorman, A.H., 2012. The Metabolic Pattern of Societies: Where Economists Fall Short. Routledge, London, UK.

Giampietro, M., Pimentel, D., 1991. Energy analysis models to study the biophysical limits for human exploitation of natural processes., in: Rossi, C., Tiezzi, E. (Eds.), Ecological Physical Chemistry. Elsevier, Amsterdam, The Netherlands, pp. 139–184.

Glansdorff, P., Prigogine, I., 1971. Thermodynamic Theory of Structure, Stability and Fluctuations. Wiley, New York, USA.

González de Molina, M., Toledo, V.M., 2014. Social metabolism: Origins, history, approaches, and main publications, in: The Social Metabolism: A Socio-Ecological Theory of Historical Change. Springer, pp. 43–58.

Gower, S.T., Krankina, O., Olson, R.J., Apps, M., Linder, S., Wang, C., 2012. NPP Boreal Forest: Consistent Worldwide Site Estimates, 1965-1995, R1. Data set. Available online [http://daac.ornl.gov]. Oak Ridge National Laboratory Distributed Active Archive Center, Oak Ridge, Tennessee, U.S.A. doi:10.3334/ORNLDAAC/611.

Grene, M., 1969. Hierarchy: one word, how many concepts, in: Whyte, L.L., Wilson, A.G., Wilson, D. (Eds.), Hierarchical Structures. American Elsevier Publishing Company, New York, USA, pp. 56–58.

Iberall, A., Soodak, H., Arensberg, C., 1980. Homeokinetic Physics of Societies, A New Discipline: Autonomous groups, Cultures, Polities, in: Real, H., Ghista, D., Rau, C. (Eds.), Perspectives in Biomechanics. Harwood Academic Press, New York, USA, pp. 433–528.

Jørgensen, S.E., 2012. Introduction to systems ecology. CRC Press/Taylor & Francis.

Karr, J.R., 1996. Ecological integrity and ecological health are not the same, in: Schulze, P.C. (Ed.), Engineering within Ecological Constraints. National Academy Press, Washington, DC, pp. 97–109.

Kay, J.J., Allen, T.F.H., Fraser, R., Luvall, J., Ulanowicz, R., 2001. Can we use energy based indicators to characterize and measure the status of ecosystems, human, disturbed and natural?, in: Ulgiati, S., Brown, M.T., Giampietro, M., Herendeen, R., Mayumi, K. (Eds.), Proceedings of the International Workshop: Advances in Energy Studies: Exploring Supplies, Constraints and Strategies. SGE Editoriali, Padova, pp. 121–133.

Kay, J.J., Regier, H., 2000. Uncertainty, complexity, and ecological integrity: Insights from an ecosystem approach., in: Crabbé, P., Holland, A., Ryszkowski, L., Westra, L. (Eds.), Implementing Ecological Integrity: Restoring Regional and Global Environmental and Human Health. Kluwer Academic Publishers, Dordrecht, The Netherlands, pp. 121–156.

Koestler, A., 1969. Beyond atomism and holism—the concept of the holon, in: Koestler, A., Smythies, J.R. (Eds.), Beyond Reductionism. Hutchinson, London, UK, pp. 192–232.

Lambers, H., Chapin III, F.S., Chapin, F.S., Pons, T.L., 2008. Plant Physiological Ecology, 2nd ed. Springer.

Lewis, S.L., Maslin, M.A., 2015. Defining the Anthropocene. Nature 519, 171–180.

Lotka, A.J., 1925. Elements of Physical Biology. Williams and Wilkins Company, Baltimore, MD, USA.

Madrid, C., Cabello, V., Giampietro, M., 2013. Water-use sustainability in socioecological systems: A multiscale integrated approach. Bioscience 63, 14–24.

Margalef, R., 1968. Perspectives in Ecological Theory. University of Chicago Press, Chicago, USA.

Margalef, R., 1963. On Certain Unifying Principles in Ecology. Am. Nat. 97, 357–374.

Maturana, H.R., Varela, F.J., 1980. Autopoiesis and Cognition: The Realization of the Living. D. Reidel Publishing Company.

Mayumi, K., 2001. The Origins of Ecological Economics: The Bioeconomics of Georgescu-Reogen. Routledge, New York, USA.

Mayumi, K., Giampietro, M., 2014. Proposing a general energy accounting scheme with indicators for responsible development: Beyond monism. Ecol. Indic. 47, 50–66.

Millennium Assessment, 2005. Ecosystems and Human Well-Being: Synthesis. Island Press, Washington D.C., USA.

Navarro, M.N. V, Jourdan, C., Sileye, T., Braconnier, S., Mialet-Serra, I., Saint-Andre, L., Dauzat, J., Nouvellon, Y., Epron, D., Bonnefond, J.M., Berbigier, P., Rouziere, A., Bouillet, J.P., Roupsard, O., 2008. Fruit development, not GPP, drives seasonal variation in NPP in a tropical palm plantation. Tree Physiol. 28, 1661–74.

Nicholls, C., Altieri, M., Vázquez, L., 2016. Agroecology: Principles for the Conversion and Redesign of Farming Systems. J. Ecosyst. Ecography 1.

Nicolis, G., Prigogine, I., 1977. Self-organization in Non-equilibrium Systems: from Dissipative Structures to Order through Fluctuations. Wiley, New York.

Obst, C.G., 2015. Reflections on natural capital accounting at the national level. Sustain. Accounting, Manag. Policy J. 6, 315–339.

Odum, E.P., 1985. Trends expected in stressed ecosystems. Bioscience 35, 419–422.

Odum, E.P., 1969. The strategy of ecosystem development. Science 164, 262–270.

Odum, H.T., 1996. Environmental Accounting: Emergy and Environmental Decision Making. Wiley, New York, USA.

Odum, H.T., 1994. Ecological and General Systems: An Introduction to Systems Ecology. University Press of Colorado, Niwot, CO, USA

Odum, H.T., 1983. Systems Ecology: An Introduction. Wiley.

Odum, H.T., 1975. Implications of energy use on environmental conservation and future ways of life, in: Thirteenth Technical Meeting of IUCN. IUCN, Kinshasa, Zaire, pp. 165–177.

Odum, H.T., 1971. Environment, Power and Society. Wiley, New York, USA.

Odum, H.T., 1957. Trophic structure and productivity in Silver Springs. Ecol. Monogr. 27, 55–112.

Odum, H.T., 1956. Primary production in flowing waters. Limnol. Oceanogr. 1, 102–117.

Odum, H.T., Odum, E.C., 1976. Energy Basis for Man and Nature. McGraw-Hill, New York, USA.

Odum, H.T., Pinkerton, R.C., 1955. Time's speed regulator: the optimum efficiency for maximum power output in physical and biological systems. Am. Sci. 43, 331–343.

Odum, W.E., Odum, E.P., Odum, H.T., 1995. Nature's pulsing paradigm. Estuaries 18, 547–555.

Padovan, D., 2000. The concept of social metabolism in classical sociology. Theomai Estud. sobre Soc. Nat. y Desarro. 2, 1–36.

Pattee, H.H. ed. edited by H.H.P., 1973. Hierarcht theorythe challenge of complex systems. George Braziller, New York, USA.

Pedersen, B., Nybø, S., Skarpaas, O., 2013. Ecological framework for the Nature Index, A more rigorous approach to the determination of reference values and selection of indicators. Trondheim, Norway.

Pimentel, D., Westra, L., Noos, R.F. (Eds.), 2000. Ecological Integrity: Integrating Environment, Conservation, and Health. Island Press, Washington D.C.

Prigogine, I., 1955. Introduction to thermodynamics of irreversible processes, 2nd ed. Interscience Publishers, New York, USA.

Reza, M.I.H., Abdullah, S.A., 2011. Regional index of ecological Integrity: A need for sustainable management of natural resources. Ecol. Indic. 11, 220–229.

Rockström, J., Steffen, W., Noone, K., Persson, Å., F. S. Chapin, I., Lambin, E., Lenton, T.M., Scheffer, M., Folke, C., Schellnhuber, H., Nykvist, B., Wit, C.A. De, Hughes, T., Leeuw, S. van der, Rodhe, H., Sörlin, S., Snyder, P.K., Costanza, R., Svedin, U., Falkenmark, M., Karlberg, L., Corell, R.W., Fabry, V.J., Hansen, J., Walker, B., Liverman, D., Richardson, K., Crutzen, P., Foley, J., 2009. Planetary Boundaries : Exploring the safe operating space for humanity. Ecol. Soc. 14, 32.

Schneider, E.D., Kay, J.J., 1994. Life as a manifestation of the 2nd law of thermodynamics. Math. Comput. Model. 19, 25–48.

Schrödinger, E., 1967. What is life? : the physical aspect of the living cell with Mind and matter. Cambridge University Press, Cambridge, UK.

Serrano-Tovar, T., Giampietro, M., 2014. Multi-scale integrated analysis of rural Laos: Studying metabolic patterns of land uses across different levels and scales. Land use policy 36, 155–170.

Simon, H., 1962. The architecture of complexity. Procedings Am. Philos. Soc. 106, 467–482.

Steffen, W., Broadgate, W., Deutsch, L., Gaffney, O., Ludwig, C., 2015a. The trajectory of the Anthropocene: The Great Acceleration. Anthr. Rev. 2, 81–98.

Steffen, W., Grinevald, J., Crutzen, P.J., McNeill, J., 2011. The Anthropocene: conceptual and historical perspectives. Philos. Trans. R. Soc. London, A 369, 842–867.

Steffen, W., Richardson, K., Rockström, J., Cornell, S.E., Fetzer, I., Bennett, E.M., Biggs, R., Carpenter, S.R., de Vries, W., de Wit, C.A., Folke, C., Gerten, D., Heinke, J., Mace, G.M., Persson, L.M., Ramanathan, V., Reyers, B., Sörlin, S., 2015b. Planetary boundaries: Guiding human development on a changing planet. Science 347, 1259855.

Ulanowicz, R., 1997. Ecology, the Ascendent Perspective. Columbia University Press, New York, USA.

Ulanowicz, R., 1986. Growth and Development: Ecosystems Phenomenology. Springer, New York, USA.

Ulanowicz, R.E., 2000. Ascendancy: A Measure of Ecosystem Performance, in: Handbook of Ecosystem Theories and Management. CRC Press, p. 303.

Ulgiati, S., Brown, M.T., 1998. Monitoring patterns of sustainability in natural and man-made ecosystems. Ecol. Modell. 108, 23–36.

UN, 2014a. System of Environmental-Economic Accounting 2012-Central Framework. United Nations Publications.

UN, 2014b. System of Environmental-Economic Accounting 2012-Experimental Ecosystem Accounting. United Nations Publications.

UN, 2014c. System of Environmental-Economic Accounting 2012-Applications and Extensions. United Nations Publications.

Vardon, M., Burnett, P., Dovers, S., 2016. The accounting push and the policy pull: balancing environment and economic decisions. Ecol. Econ. 124, 145–152.

Wackernagel, M., Rees, W., 1998. Our ecological footprint: reducing human impact on the earth. New Society Publishers, Gabriola Island, British Columbia, Canada.

Waltner-Toews, D., Kay, J.J., Lister, N.M.E. (Eds.), 2008. The Ecosystem Approach: Complexity, Uncertainty, and Managing for Sustainability. Columbia University Press, New York, USA.

White, L.A., 1943. Energy and the evolution of culture. Am. Anthropol. 45, 335–356.

World Bank, 2010. The Changing Wealth of Nations. The World Bank, Washington D.C., USA.

Xie, X., Wang, Q., Dai, L., Su, D., Wang, X., Qi, G., Ye, Y., 2011. Application of China's National Forest Continuous Inventory Database. Environ. Manage. 48, 1095–1106.

Zalasiewicz, J., Williams, M., Smith, A., Barry, T.L., Coe, A.L., Bown, P.R., Brenchley, P., Cantrill, D., Gale, A., Gibbard, P., 2008. Are we now living in the Anthropocene? Gsa Today 18, 4.

Zampella, R.A., Bunnell, J.F., Laidig, K.J., Procopio, N.A., 2006. Using multiple indicators to evaluate the ecological integrity of a coastal plain stream system. Ecol. Indic. 6, 644–663.

Zhao, M., Zhou, G.-S., 2005. Estimation of biomass and net primary productivity of major planted forests in China based on forest inventory data. For. Ecol. Manage. 207, 295–313.

Zipf, G., 1941. National Unity and Disunity: The Nation as a Bio-social Organism. Principia Press, Bloomington.

# FIGURE CAPTION

Figure 1. Energy flow diagram for a generic ecosystem, showing the different energy forms interacting between them and with the environment through the environmental window of attention (grey line). Inside and outside views of the different levels are shown.

Figure 2. Hierarchical representation of ecosystem and societal levels associated to biomass production and consumption, respective. The graph shows the link between the two non-equivalent descriptive domains (inside and outside views) and the reinforcement loops. After Madrid et al. (2013).

Figure 3. Flows of energy and materials associated with the reproduction of biomass as a fund, and human alteration of this self-organization process by withdrawal and human control (inputs and decrease of competitors).

Figure 4. Flow/fund ratio (negentropic costs - vertical axis) and fund size (standing biomass -horizontal axis) for tropical forests and selected crops cultivated under tropical conditions.

Figure 5. Flow/fund ratio (negentropic costs - vertical axis) and fund size (standing biomass -horizontal axis) for boreal forests and selected tree plantations under boreal conditions.

Figure 6. Example of integrated assessment for different land use scenarios about tree plantations under boreal conditions. For each scenario, it is presented the share of the different land uses and the negentropic costs associated.